\documentclass[twocolumn,showpacs,amsmath,amssymb,nofootinbib,amsfonts,showkeys]{revtex4}
\usepackage{graphicx}
\usepackage{bm}

\begin{document}

\title{Evolution of scalar perturbations in cosmology with quintessential dark energy}
\author{B. Novosyadlyj}
 \email{novos@astro.franko.lviv.ua}
\author{O. Sergijenko}
 \email{olka@astro.franko.lviv.ua}
\affiliation{Astronomical Observatory of 
Ivan Franko National University of Lviv, Kyryla i Methodia str., 8, Lviv, 79005, Ukraine}

\date{\today}

\begin{abstract}

The dynamics of expansion of the Universe and evolution of scalar perturbations are discussed for the quintessential scalar fields $Q$ with the classical Lagrangian $L=\frac{1}{2}Q_{;i}Q^{;i}-U(Q)$ satisfying the additional condition $w=const$ or $c^2_a=0$. Both quintessential fields are studied for the same cosmological model. It is shown that the accelerated expansion of the Universe is caused by the effect of rolling down of the field to minimum. At the early epoch the contribution to dynamics of the quintessence with $w=const$ is negligible (like that of cosmological constant) while quintessence with $c^2_a=0$ mimics dust matter. In future the scalar field with $c^2_a=0$ will mimic cosmological constant. 

The systems of evolution equations for gauge-invariant perturbations of metric, matter and quintessence have been analysed analyticaly for the early stage of the Universe life and numerically up to the present epoch. It is shown that amplitudes of the adiabatic matter density perturbations grow similarly in both models (and like in $\Lambda$CDM-model), but time dependences of different amplitudes of the quintessence perturbations are varied: gauge-invariant variables $D_g^{(Q)}$ and $D_s^{(Q)}$ decay from initial constant value after the particle horizon entry while $D^{(Q)}$ and $V^{(Q)}$ grow at the early stage before the horizon entry and decay after that -- in the quintessence-dominated epoch, when gravitational potential starts to decay -- so, that at the current epoch they are approximately two orders lower than matter ones on the supercluster scales. Therefore, on the subhorizon scales the quintessential scalar fields are smoothed out while the matter clusters.

It is also shown that both quintessential scalar fields suppress the growth of matter density perturbations and the amplitude of gravitational potential. In these QCDM-models -- unlike $\Lambda$CDM ones -- such suppression is scale dependent and more visible for the quintessence with $c^2_a=0$. 

\end{abstract}



\maketitle

\section{Introduction}
Cosmological observations of the last decade surely assert that the main part of the energy density of the Universe -- more than 70\% --  belongs to the unknown essence, called ''dark energy''. Its cosmological mission is to provide the accelerated expansion of the Universe, revealed from exploration of SN Ia's in the distant galaxies and temperature fluctuation power spectrum of cosmic microwave background. The cosmological $\Lambda$CDM-model, based on the Einstein equations with cosmological constant (see \cite{apunevych2007,spergel2007,komatsu2008} and references therein), describes very well almost whole set of the observational data on dynamics of expansion of the Universe and formation of its large-scale structure. But physical interpretation of the cosmological constant is rather problematic \cite{peebles1993,sahni2000,carrol2001,peebles2003,bousso2008,shapiro2008}. Therefore alternative approaches  --
new physical fields (classical scalar field -- quintessence, tachyon field, k-essence, phantom field, quintom field), Chaplygin gas,
gravity and general relativity modifications, multidimensional gravity, branes and others -- are intensively analysed (see reviews 
\cite{sahni2000,carrol2001,padmanabhan2003,peebles2003,copeland2006,bludman2007,frieman2008} and special issue of Gen. Relativ. Gravit., 2008, v.40) now. Up to now none of them has crucial preferability from observational or theoretical point of view. Therefore each of them must be comprehensively studied. Here we restrict ourselves to quintessential scalar fields with classical Lagrangian $L=Q_{;i}Q^{;i}/2-U(Q)$ in the dark energy -- matter dominated Universe.

The quintessence model can be defined by setting of the appropriate potential $U(Q)$ or equation of state (EoS) parameter $w_{Q}\equiv p_{Q}/c^2\rho_{Q}$. There is a dozen or more physically-motivated shapes of the potential $U(Q)$: exponential, double exponential, exponential with inverse power, power-law, etc. The dynamics of such scalar fields is intensively studied (see review \cite{copeland2006}). The EoS parameter of dark energy completely defines the background dynamics as well as the evolution of cosmological perturbations \cite{hu1998,hu1999,linder2008}. Since observational data on SN Ia magnitude -- redshift relation and cosmic microwave anisotropy give relatively narrow ranges of dark energy density and EoS parameter values, it looks quite attractive to establish the potential $U(Q)$ using these data and analyse the background dynamics and perturbative properties of such scalar field which are not studied widely enough.

In our previous papers we have constructed the potentials of scalar fields with classical and tachyonic Lagrangian leading to the constant EoS parameter $w_{Q}=const$ \cite{sergijenko2008a} and analysed the background dynamics and perturbative properties of such scalar fields \cite{sergijenko2008b}. It was shown that cosmological model with cold dark matter and such types of the scalar field ($QCDM$-model) agrees slightly better with the accessible today observable data than the $\Lambda$CDM-model. But difference of quantitative merits of goodness is not large enough to pick out one of them at confidential level of  $1\sigma$. Since the degeneracies between model parameters of dark energy and cosmological parameters \cite{kunz2007,durrer2008,hlozek2008,linder2008,zhdanov2008} exist for the background dynamics, the complete analysis of linear density perturbations in both dark matter and dark energy components is important for improvement of dark energy observational tests. Among the large number of free quintessence parameters and unknown initial values of quintessence perturbation modes there is only small part of models, for which the evolution of perturbations has been studied. 
The general conclusion is that magnitudes of dark energy density perturbations on scales smaller than horizon are essentially lower than corresponding magnitudes of matter density ones. But character of their evolution depends strongly on the scalar field model (its potential, time variation of EoS parameter, sound speed, etc.), initial conditions, scale of perturbations and gauge (see for example  \cite{dave2002,doran2003,dedeo2003,bean2004,bartolo2004,liu2004,kulinich2004,dutta2007,unnikrishnan2008}).

Here the special attention should be paid to the EoS parameter of dark energy $w_{Q}$, which can be constant or varying in time. The temporal variation of the dark energy EoS parameter is often presented by linear fitting formula with two \cite{chevallier2001} or three \cite{komatsu2008} parameters to be estimated. Other functional dependences of $w_Q$ on scale factor or redshift can be found in \cite{copeland2006,linder2006,sahni2006}.
Here we study the parametrization of the equation of state, which needs only 1 additional quantity with clear physical meaning -- the adiabatic speed of sound $c_a^2\equiv \dot{p}_Q/c^2\dot{\rho}_Q$ (the analysis of generalized dark sector components can be found in the early works \cite{hu1998,hu1999}). In general, $c_a^2$ is the unknown function of time. However, taking into account the simplicity we restrict ourselves to $c_a^2=const$, so it is regarded only as the second physical parameter defining the equation of state of dark energy (the first one -- the present value of $w_Q$). 

In this paper we undertake the comparative analysis the evolution of gauge-invariant variables of the scalar perturbations in the model with non-relativistic matter ($p_{M}\ll c^2\rho_{M}$) and scalar field which we define by classical Lagrangian with potential constructed for two cases ($w_{Q}=const$ and $c^2_a=0$) in the concordance cosmological models. These cases have been chosen because they allow us to obtain analytical solutions, which seems to look very attractive in the world of numerical computations. We assume the adiabatic initial conditions for matter and dark energy scalar perturbations.

\section{Background cosmological and scalar field models}

We consider the homogeneous and isotropic flat Universe with metric of 4-space 
\begin{eqnarray*}
&&ds^2=g_{ij} dx^i dx^j =c^2dt^2-a^2(t)\delta_{\alpha\beta} dx^{\alpha}dx^{\beta}\\
&&=a^2(\eta)(d\eta^2-\delta_{\alpha\beta} dx^{\alpha}dx^{\beta}),
\end{eqnarray*}
where the factor $a(t)$ is the scale factor, normalized  to 1 at  the current epoch $t_0$, $\eta$ is conformal time ($cdt=a(\eta)d\eta$). Henceforth we also put $c=1$, so the time variable $t\equiv x_0$  has the dimension of a length. Here and below the latin indices $i,\,j,\,...$ run from 0 to 3, the greek ones -- over the spatial part of the metric: $\nu,\, \mu,\,...$=1, 2, 3. 

If the Universe is filled with non-relativistic matter (cold dark matter and baryons) and quintessence which interact only gravitationally (minimal coupling) then the dynamics of its expansion is completely described by the Einstein equations 
\begin{equation}
R_{ij}-{\frac{1}{2}}g_{ij}R=8\pi G \left(T_{ij}^{(M)}+T_{ij}^{(Q)}\right),
\label{Einstein-eq}
\end{equation}
where $R_{ij}$ is the Ricci tensor and $T_{ij}^{(M)}$, $T_{ij}^{(Q)}$ -- energy-momentum tensors of Matter $(M)$ and Quintessence $(Q)$. If these components interact only gravitationally then each of them satisfy the differential energy-momentum conservation law separately:
\begin{equation}
T^{i\;\;(M,Q)}_{j\;;i}=0
\label{conserv-eq}
\end{equation}
(here and below ``;'' denotes the covariant derivative with respect to the coordinate with given index in the space with metric $g_{ij}$). For the perfect fluid with density $\rho_{(M,Q)}$ and pressure $p_{(M,Q)}$, related by the equation of state $p_{(M,Q)}=w_{(M,Q)}\rho_{(M,Q)}$, it gives
\begin{equation}
\dot{\rho}_{(M,Q)}=-3\frac{\dot a}{a} \rho_{(M,Q)}(1+w_{(M,Q)}) \label{eqconsm}
\end{equation}
(here and below a dot over the variable denotes the derivative with respect to the conformal time: ``$\dot{\;\;}$''$\equiv d/d\eta$). The matter is considered to be non-relativistic, so  $w_M^{ }=0$ and $\rho_M^{ }=\rho_M^{(0)}a^{-3}$ (here and below ``0'' denotes the present values). 

We assume the quintessence to be a scalar field $Q({\bf x},\eta)$ with classical Lagrangian 
\begin{equation}
L=\frac{1}{2}Q_{;i}Q^{;i}-U(Q), \label{lagr_cf}
\end{equation}
where $U(Q)$ is the field potential.
We suppose also the background scalar field to be homogeneous ($Q({\bf x},\eta)=Q(\eta)$), so its energy density and pressure depend only on time:
\begin{eqnarray}
\rho_{Q}(\eta)=\frac{1}{2a^2}\dot{Q}^{2}+U(Q),\,\,\,\,\,
p_{Q}(\eta)=\frac{1}{2a^2}\dot{Q}^{2}-U(Q).\label{rho_q}
\end{eqnarray}
Then the conservation law (\ref{conserv-eq}) gives the scalar field evolution equation (called the Klein-Gordon one)
\begin{eqnarray}
 \ddot{Q}+2aH\dot{Q}+a^2\frac{dU}{dQ}=0, \nonumber
\end{eqnarray}
where $H=\dot{a}/{a^2}$ is the Hubble parameter for any moment of conformal time $\eta$.

\begin{figure}
\includegraphics[width=8cm]{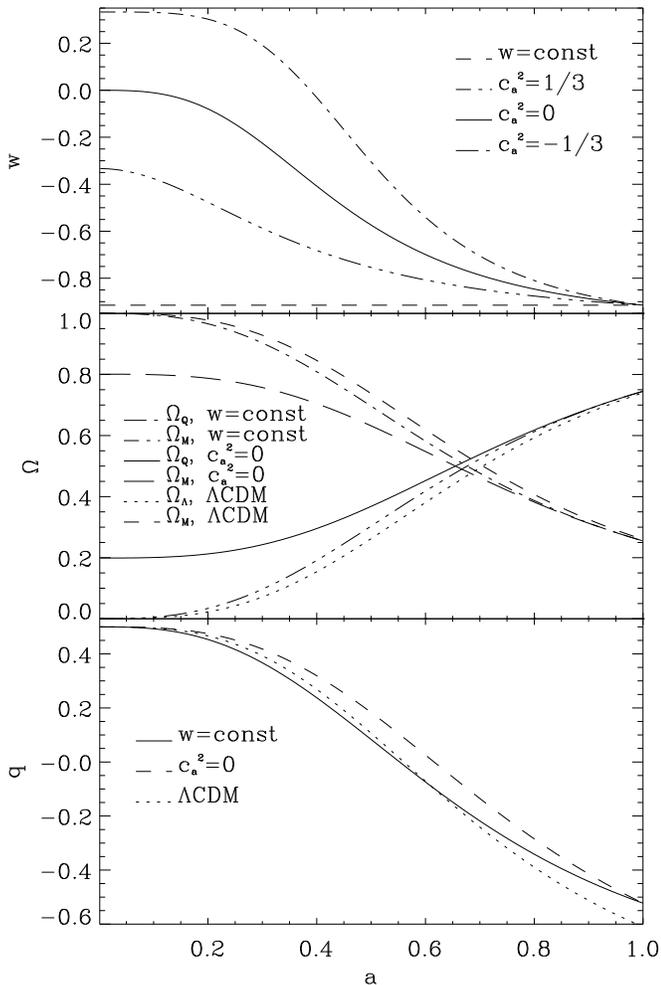}
\caption{Top: the dependence of EoS parameter $w$ on scale factor $a$ for $c^2_a=1/3,\,\,0,\,\,-1/3$ and $w=c^2_a=const$. Middle: the dynamics of expansion of the homogeneous Universe in the model with non-relativistic matter and quintessential scalar field  with $c^2_a=0$ and $w=c^2_a=const$ with best fitting cosmological parameters from Spergel et al. (2007) ($\Omega_{Q}=0.745$, $w=-0.915$, $\Omega_M=0.255$, $h=0.7$): matter and quintessence densities in units of the critical one. Bottom: the evolution of acceleration parameter. For comparison we show also the corresponding dependences for $\Lambda$CDM-model with  $\Omega_{Q}=0.74$, $\Omega_M=0.26$ and $h=0.73$ (Spergel et al. (2007), Apunevych et al. (2007)). } 
\label{fig1}
\end{figure}

We specify the model of quintessence using two thermodynamical parameters: the EoS parameter $w_{Q}\equiv p_Q/\rho_Q$ and the adiabatic speed of sound
$c^2_a\equiv \dot{p}_Q/\dot{\rho}_Q$. In general case they are connected by equation
$$\frac{dw/d\ln{a}}{3(1+w)}=w-c^2_a$$
(here and below we omit index $Q$ for $w_{Q}$).
If the time dependence of $w$ is known then $c_a^2$ is defined unambiguously, if $c_a^2$ is determined then the initial value $w_0$ must be defined additionaly, so, the EoS parameter has 2 degrees of freedom: a function and a constant. For other parametrizations see \cite{copeland2006,linder2006,sahni2006}.
Since the constraints for time dependence of $w$ or $c^2_a$ are not established well we consider
two simple cases: $w=const$  and $c^2_a=const$. In the first case $c^2_a=w$ and in the second one  
$1+w(a)=(1+c^2_a)(1+w_0)/\left(1+w_0-(w_0-c^2_a)a^{3(1+c^2_a)}\right).$
This equation has obvious asymptotical behaviour: when $a\rightarrow0$ $w \rightarrow c^2_a$ and when $a\rightarrow\infty$ $w \rightarrow -1$. So, at early epoch the dark energy mimics dust matter ($w\approx 0$) for $c^2_a=0$ or radiation ($w\approx 1/3$) for $c^2_a=1/3$. In future such scalar field will mimic cosmological constant ($w\approx -1$). The time dependences of EoS parameter for both cases are shown in Fig.\ref{fig1}. The equation (\ref{eqconsm}) has the analytical solutions for two cases:
\begin{itemize}
 \item $w=const$: $\rho_{Q}(a)=\rho_{Q}^{(0)}a^{-3(1+w)}$ and
\item $c^2_a=0$: $\rho_{Q}(a)=\rho_{Q}^{(0)}\left[(1+w_0)a^{-3}-w_0\right]$,
\end{itemize}
so it's possible to simplify formulae and calculations and we will analyse only this two cases now.

If the parametrization of EoS parameter is given, it is possible to apply reverse engineering and construct the fields $Q$ and potentials $U(Q)$. From (\ref{rho_q}) one simply obtains:
\begin{eqnarray*}
Q(a)-Q_0=\pm\int_1^a\frac{\sqrt{\rho_Q(1+w)}}{aH},\,\,\,U(a)=\frac{\rho_Q(1-w)}{2}.
\end{eqnarray*}

\noindent If the integral for $Q$ can be expressed via functions that could be inverted to obtain $a(Q-Q_0)$, then $U(Q-Q_0)$ can be easily written in analytical form.

So, from the Einstein and field equations we deduce the time dependences of the Hubble $H$ and acceleration $q$ parameters as well as the evolution of the scalar field $Q$ and potential $U(Q)$:
\begin{widetext}
\begin{eqnarray}
&&H=H_0a^{-\frac{3}{2}}\sqrt{1-\Omega_{Q}+\Omega_{Q}a^{-3w}},\hskip1.6cm
q=\frac{1}{2}\frac{1-\Omega_{Q} +(1+3w)\Omega_{Q}a^{-3w}}
{1-\Omega_{Q}+\Omega_{Q}a^{-3w}},\label{q}\\
&&Q(a)-Q_{0}=\pm\frac{1}{2\sqrt{6\pi G}}\frac{\sqrt{1+w}}{w}
\ln\left(\frac{\sqrt{(1-\Omega_{Q})a^{3w}+\Omega_{Q}}-\sqrt{\Omega_{Q}}}{\sqrt{(1-\Omega_{Q})a^{3w}+\Omega_{Q}}+\sqrt{\Omega_{Q}}}\frac{1+\sqrt{\Omega_{Q}}}{1-\sqrt{\Omega_{Q}}}\right),\\
&&U(Q-Q_{0})=\frac{3H_{0}^{2}}{8\pi G}\Omega_{Q}\frac{1-w}{2}\left[ch\left(\sqrt{6\pi G}(Q-Q_{0})\frac{w}{\sqrt{1+w}}\right)\mp\frac{1}{\sqrt{\Omega_{Q}}}sh\left(\sqrt{6\pi G}(Q-Q_{0})\frac{w}{\sqrt{1+w}}\right)\right]^{2\frac{1+w}{w}}
\end{eqnarray}
\end{widetext}
for $w=const$ and 
\begin{widetext}
\begin{eqnarray}
&&H=H_0a^{-\frac{3}{2}}\sqrt{1+\Omega_{Q}w_0-\Omega_{Q}w_0a^3},\hskip1.6cm
q=\frac{1}{2}\frac{1+w_0\Omega_{Q}+2w_0\Omega_{Q}a^3}
{1+\Omega_{Q}w_0-\Omega_{Q}w_0a^3},\label{q0}\\
&&Q(a)-Q_{0}=\pm\frac{1}{2\sqrt{6\pi G}}\sqrt{\frac{\Omega_Q(1+w_0)}{1+\Omega_Qw_0}}
\ln\left(\frac{\sqrt{1+\Omega_Qw_0(1-a^3)}-\sqrt{1+\Omega_Qw_0}}{\sqrt{1+\Omega_Qw_0(1-a^3)}+\sqrt{1+\Omega_Qw_0}}\frac{1+\sqrt{1+\Omega_{Q}w_0}}{1-\sqrt{1+\Omega_Qw_0}}\right),\\
&&U(Q-Q_0)=\frac{3H_0^2}{8\pi G}\frac{\Omega_Q(1+w_0)}{2}\left[ch\left(\sqrt{6\pi G}(Q-Q_{0})\sqrt{\frac{1+\Omega_{Q}w_0}{\Omega_Q(1+w_0)}}\right)\right.\nonumber 
\end{eqnarray}
\end{widetext}
\begin{widetext}
\begin{eqnarray}
&&\left.\mp\frac{1}{\sqrt{1+\Omega_{Q}w_0}}sh\left(\sqrt{6\pi G}(Q-Q_{0})\sqrt{\frac{1+\Omega_{Q}w_0}{\Omega_Q(1+w_0)}}\right)\right]^2-\frac{3H_0^2}{8\pi G}\Omega_Qw_0
\end{eqnarray}
\end{widetext}
for $c^2_a=0$.
For both models there are 2 independent solutions for the field (the growing one corresponds to sign ``+'' and the decaying one to sign ``-'') and 2 symmetrical with respect to $Q-Q_0$ potentials exist. However, the physical consequences of both these solutions are the same \cite{sergijenko2008a}, so from now we restrict ourselves only to the growing one. The variety of scalar field potentials was presented in \cite{copeland2006}. The potential for $w=const$ can be also found in \cite{sahni2003,sergijenko2008a}, the potential for $c_a^2=0$ belongs generally to the family of double exponential potentials (with additional constant term), but both they differ from the physically-motivated ones, for which the evolution of scalar linear perturbations was studied by other authors.

We must note that the asymptotic behaviour at $a\rightarrow 0$ of the expansion rate $H(a)$ and acceleration parameter $q(a)$ in both cases is the same and similar to that in $\Lambda$CDM-model: $H\propto a^{-3/2}$, $q\rightarrow 1/2$. At current epoch the parameters of expansion dynamics are the same ($H_0$ and $q_0=(1+3w\Omega_{Q})/2$) for both models. But their asymptotic behaviour at $a\rightarrow \infty$ is different: in $w=const$ quintessence $H\rightarrow H_0a^{-\frac{3}{2}(1+w)}\sqrt{\Omega_Q}$, $q\rightarrow (1+3w)/2$, $\rho_Q\rightarrow0$ and in $c^2_a=0$ quintessence $H\rightarrow H_0\sqrt{-w_0\Omega_Q}$, $q\rightarrow -1$, $\rho_Q\rightarrow-w_0\rho_Q^{(0)}$. The energy densities of both fields evolve similarly but have different asymptotic regimes: in the quintessence with $w=const$ $\rho_Q/\rho_M=\Omega_Qa^{-3w}/(1-\Omega_Q)$ always while in the $c^2_a=0$ quintessence at $a\rightarrow 0$  $\rho_Q/\rho_M\rightarrow (1+w_0)\Omega_Q/(1-\Omega_Q)$ and at $a\rightarrow \infty$ $\rho_Q/\rho_M\rightarrow -w_0\Omega_Qa^3/(1-\Omega_Q)$. So, the scalar field with $c^2_a=0$ behaves as cold dark matter at the early epoch and will mimic the cosmological constant in far future.

The different asymptotic behaviour of these fields is caused by their intrinsic properties. In the $w=const$ quintessence the negative pressure stiffly follows its energy density and their relation is always constant. In the $c^2_a=0$ quintessence the negative pressure is always constant: $p_Q=3H_0^2w_0\Omega_Q/8\pi G$. So, it is insignificant in the early epoch when $a\rightarrow 0$ and $\rho_Q\rightarrow \infty$ for the model of the Universe filled only with dust matter and quintessential dark energy, and important in the late one when $w\rightarrow -1$.  

The dynamics of expansion of homogeneous Universe in the model with non-relativistic matter and quintessential scalar field  with $c^2_a=0$ and $w=c^2_a=const$ is shown in Fig.\ref{fig1}. For both models we assume best fitting cosmological parameters from \cite{spergel2007} ($\Omega_{Q}=0.745$, $w=w_0=-0.915$, $\Omega_M=0.255$, $h=0.7$). For comparison we show also the corresponding dependences in $\Lambda$CDM-model with  $\Omega_{\Lambda}=0.74$, $\Omega_M=0.26$ and $h=0.73$ \cite{spergel2007,apunevych2007}. 

\begin{figure*} 
\includegraphics[width=16cm]{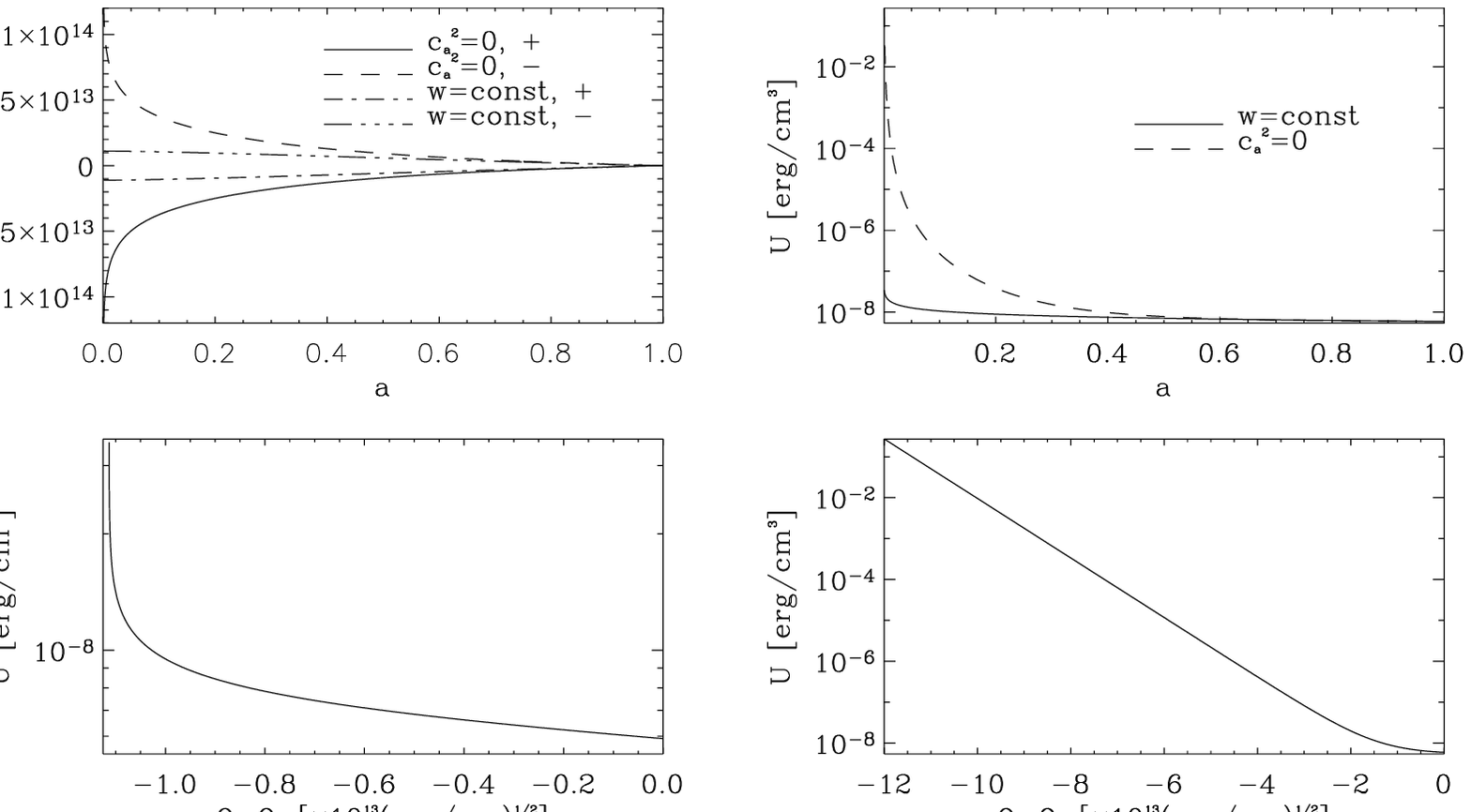}
\caption{Top: the evolution of fields $Q(a)$ (left), potentials $U(a)$ (right) for quintessence with $w=const$  and $c^2_a=0$.  Bottom: rolling  down of the fields $Q$ to the minimum $U(Q)=0$ which is located at $Q\rightarrow\infty$  ($a\rightarrow\infty$) for cases of $w=const$ (left) and $c^2_a=0$ (right).}
\label{fig3}
\end{figure*}

We have constructed the potentials of quintessential scalar fields with  $w=const$ \cite{sergijenko2008a} and $c^2_a=0$ for QCDM cosmological model with best fitting parameters obtained from WMAP and SNIa data \cite{spergel2007}. The evolution of fields $Q(a)$, potentials $U(a)$ and rolling down of the fields $Q$ to the minimum which is located at $Q\rightarrow\infty$  ($a\rightarrow\infty$) are shown in Fig.\ref{fig3}. The discussion of influence of parameter determination uncertainties on potential of field with $w=const$ can be found in \cite{sergijenko2008a}.

So, the difference of the homogeneous Universe expansion dynamics in $\Lambda$CDM- and such QCDM-models is too small to discriminate them using the avialable datasetets. That's why in the next sections we will analyse the linear stage of growth of scalar perturbations of matter and dark energy. For this we will use gauge-invariant approach developed by \cite{bardeen1980,kodama1984,durrer2001}.

\section{Evolution of scalar perturbations}

For analysis of the scalar linear perturbations the conformal-Newtonian gauge with space-time metric  
\begin{eqnarray}
&&ds^2=a^2(\eta)[(1+2\Psi(\textbf{x},\eta))d\eta^2\nonumber\\
&&-(1+2\Phi(\textbf{x},\eta))\delta_{\alpha\beta}dx^{\alpha}dx^{\beta}]
\end{eqnarray}
is convenient. Here $\Psi(\textbf{x},\eta)$ and $\Phi(\textbf{x},\eta)$ are gauge-invariant perturbations of metric \cite{bardeen1980} called Bardeen's potentials. If proper anisotropy of medium equals zero then $\Psi(\textbf{x},\eta)=-\Phi(\textbf{x},\eta)$. Dust matter and scalar fields have such property \cite{kodama1984}. In the linear perturbation theory the Fourier decomposition is used, so spatial
dependences of all variables can be substituded by corresponding Fourier amplitudes. For example,   $\Psi(\textbf{x},\eta)\rightarrow \Psi(k,\eta)$, where $k$ is wave number.  Henceforth, saying about metric $\Psi(\textbf{x},\eta)$, matter density $\delta^{(M)}(\textbf{x},\eta)\equiv (\rho_M(\textbf{x},\eta)- \bar{\rho}_M(\eta))/\bar{\rho}_M(\eta)$, its peculiar velocity $V^{(M)}(\textbf{x},\eta)$, scalar field $\delta{Q}(\textbf{x},\eta)\equiv Q(\textbf{x},\eta)- \bar{Q}(\eta)$,  its energy density perturbations $\delta^{(Q)}(\textbf{x},\eta)\equiv (\rho_Q(\textbf{x},\eta)- \bar{\rho}_Q(\eta))/\bar{\rho}_Q(\eta)$ etc we mean their Fourier amplitudes $\Psi(k,\eta)$, $\delta^{(M)}(k,\eta)$, $V^{(M)}(k,\eta)$, $\delta{Q}(k,\eta)$, $\delta^{(Q)}(k,\eta)$, etc.
The metric ($\Psi(k,\eta)$), matter density and velocity perturbations ($\delta^{(M)}(k,\eta)$, 
$V^{(M)}(k,\eta)$) as well as scalar field  perturbations ($\delta{Q}(k,\eta)$,  $\delta^{(Q)}(k,\eta)$, $V^{(Q)}(k,\eta)$) in the conformal-Newtonian gauge are gauge-invariant variables \cite{kodama1984}. The energy density and velocity perturbations of quintessence, $\delta^{(Q)}$ and $V^{(Q)}$, are connected with the perturbation of field variable $\delta{Q}$ in following way:
\begin{eqnarray*}
&&\delta^{(Q)}=(1+w)\left(\frac{\dot{\delta{Q}}}{\dot{Q}}-\Psi+\frac{a^2\delta{Q}}{\dot{Q}^2}\frac{dU}{dQ}\right),\\
&&V^{(Q)}=\frac{k\delta{Q}}{\dot{Q}}.
\label{dQ}
\end{eqnarray*}
Other non-vanishing gauge-invariant perturbations of scalar field are isotropic pressure perturbation 
\begin{equation}
\pi_L^{(Q)}=\frac{1+w}{w}\left(\frac{\dot{\delta{Q}}}{\dot{Q}}-\Psi-\frac{a^2\delta{Q}}{\dot{Q}^2}\frac{dU}{dQ}\right)\nonumber
\end{equation}
and intrinsic entropy
\begin{equation}
\Gamma^{(Q)}=\pi_L^{(Q)}-\frac{c^2_a}{w}\delta^{(Q)}.\nonumber
\end{equation}
The density perturbation of any component in the conformal-Newtonian gauge $D_s\equiv\delta$, which is gauge-invariant variable, is related to the other
gauge-invariant variables of density perturbations $D$ and $D_g$ as:
\begin{equation}
  D=D_g+3(1+w)\left(\Psi+\frac{\dot{a}}{a}\frac{V}{k}\right)=D_s+3(1+w)\frac{\dot{a}}{a}\frac{V}{k},\label{ddgds}
\end{equation}
where $D_s$, $D$, $D_g$ and $V$ correspond to either $M$- or $Q$-component. Here $D_g$ is the density perturbation in the rest frame in which the fluctuations of the curvature scalar of the constant time hypersurface vanish and $D$ corresponds to the rest frame in which the 4-velocity is orthogonal to constant time hypersurface \cite{kodama1984}.

The intrinsic entropy of quintessence $\Gamma^{(Q)}$ can be presented via gauge-invariant $Q$-perturbations as follows: 
\begin{equation}
 w\Gamma^{(Q)}=(1-c^2_a)D^{(Q)}.
\label{Gamma}
\end{equation}
This equation shows that the intrinsic entropy for scalar perturbations of quintessence with $c^2_a\neq1$ is non-zero when proper energy density perturbation $D^{(Q)}$ (measured in synchronous comoving gauge) of quintessence is non-vanishing. In the first case ($w=const$) $w\Gamma^{(Q)}=(1-w)D^{(Q)}$, in the second one ($c_a^2=0$) $w\Gamma^{(Q)}=D^{(Q)}$. In the case of perturbed quintessence dissipative processes generate entropic perturbations, so we have the sound speed $c^2_s$ defined by more general relation: $c^2_s\equiv \delta p_Q/\delta \rho_Q$. The intrinsic entropy perturbation can be presented in the form: $w\Gamma^{(Q)}\equiv (c^2_s-c^2_a)D^{(Q)}$ \cite{kodama1984}. For the scalar fields with classical Lagrangian $c^2_s=1$ \cite{erickson2002,weller2003}.

\subsection{Evolution equations}

Evolution equation for scalar field perturbation $\delta{Q}(k,\eta)$ can be obtained either from Lagrange-Euler equation or from energy-momentum conservation law ${T^i_{0;i}}^{(Q)}=0$:
 \begin{eqnarray}
&&\ddot{\delta{Q}}+2aH\dot{\delta{Q}}+\left(k^2+a^2\frac{d^2U}{dQ^2}\right)\delta{Q}+2a^2\frac{dU}{dQ}\Psi\nonumber\\
&&-4\dot{\Psi}\dot{Q}=0.\label{phi_evol}
\end{eqnarray}
Thus, evolution of quintessence perturbation depends on field model ($U(Q)$), gravitational potential $\Psi$, expansion rate of the Universe $H$ and scale of perturbation $k$. 

The linearised Einstein equations for gauge-invariant perturbations of metric and energy-momentum tensor components are
\begin{widetext}
\begin{eqnarray}
&&{D_g'}^{(Q)}+\frac{3}{a}(1-w)D_g^{(Q)}+\left(\frac{k}{a^2H}+9\frac{(1-c_a^2)H}{k}\right)(1+w)V^{(Q)}+9(1+w)(1-c_a^2)\frac{\Psi}{a}=0,\label{edgQw}\\
&&{V'}^{(Q)}-\frac{2}{a}V^{(Q)}-4\frac{k\Psi}{a^2H}-\frac{k}{a^2H}\frac{D_g^{(Q)}}{1+w}=0,\label{evQw}\\
&& \Psi'+\frac{\Psi}{a}-\frac{4\pi G}{H}\left(\bar{\rho}_M\frac{V^{(M)}}{k}+\bar{\rho}_Q(1+w)\frac{V^{(Q)}}{k}\right)=0.\label{epsiw}
\end{eqnarray}
\end{widetext}
Here and below a prime denotes the derivative with respect to the scale factor $a$.
The conservation equations for matter density and velocity perturbations $\delta {T^i_{j;i}}^{(M)}=0$ in terms of the gauge-invariant variables $D_g^{(M)}$ and $V^{(M)}$ are following:
\begin{eqnarray}
&&{D_g'}^{(M)}+\frac{kV^{(M)}}{a^2H}=0,\label{cons_M1}\\
&&V'^{(M)}+\frac{V^{(M)}}{a}-\frac{k\Psi}{a^2H}=0\label{cons_M2}.
\end{eqnarray}
They are connected with the dark energy ones only via $\Psi$ and are the same for both models of quintessence. 

So, in each case we have the system of 5 first-order ordinary differential equations for 5 unknown functions $\Psi(k,a)$, $D_g^{(M)}(k,a)$, $V^{(M)}(k,a)$, $D_g^{(Q)}(k,a)$ and $V^{(Q)}(k,a)$.
From this systems of equations it is easy to obtain the systems of 2 second-order ordinary differential equations for 2 unknown functions $\Psi(k,a)$ and $\delta Q(k,a)$:
\begin{widetext}
\begin{eqnarray}
&&\Psi''+\left(\frac{7}{2}-\frac{3}{2}w\Omega_Qa^{-3(1+w)}\frac{H^2_0}{H^2}\right)\frac{\Psi'}{a}+\frac{3}{2}(1-w)\Omega_Qa^{-3(1+w)}\frac{H^2_0}{H^2}\frac{\Psi}{a^2}\nonumber\\
&&\hskip0.5cm  -a^{-\frac{3}{2}(1+w)}\frac{H_0}{H}\sqrt{6\pi G\Omega_Q(1+w)}\frac{2a\delta{Q}'+3(1-w)\delta{Q}}{2a^2}=0,\label{e2psiw}
\end{eqnarray}
\end{widetext}
\begin{widetext}
\begin{eqnarray}
&&\delta{Q}''+\left(\frac{5}{2}-\frac{3}{2}w\Omega_Qa^{-3(1+w)}\frac{H^2_0}{H^2}\right)\frac{\delta{Q}'}{a}+\left(\frac{k^2}{a^4H^2}+
\frac{9(1-w)(2+w)}{4a^2}+\frac{9w(1-w)}{4a^2}\Omega_Qa^{-3(1+w)}\frac{H^2_0}{H^2}\right)\delta{Q}\nonumber\\ 
&&\hskip0.7cm  -a^{-\frac{3}{2}(1+w)}\frac{H_0}{H}\sqrt{\frac{3}{8\pi G}\Omega_Q(1+w)}\frac{4a\Psi'+3(1-w)\Psi}{a^2}=0 \label{e2phiw}
\end{eqnarray}
\end{widetext}
for $w=const$ or 
\begin{widetext}
\begin{eqnarray}
&&\Psi''+\left(\frac{7}{2}-\frac{3}{2}w_0\Omega_Q\frac{H^2_0}{H^2}\right)\frac{\Psi'}{a}+\frac{3}{2}(1+w_0-2w_0a^3)\Omega_Qa^{-3}\frac{H^2_0}{H^2}\frac{\Psi}{a^2}-a^{-\frac{3}{2}}\frac{H_0}{H}\sqrt{6\pi G\Omega_Q(1+w_0)}\nonumber\\
&&\times\frac{2a\delta{Q}'+3\delta{Q}}{2a^2}=0,\label{e2psi0}\\
&&\delta{Q}''+\left(\frac{5}{2}-\frac{3}{2}w_0\Omega_Q\frac{H^2_0}{H^2}\right)\frac{\delta{Q}'}{a}+\left(\frac{k^2}{a^4H^2}+ 
\frac{9}{2a^2}+\frac{9}{4a^2}w_0\Omega_Q\frac{H^2_0}{H^2}\right)\delta{Q}-a^{-\frac{3}{2}}\frac{H_0}{H}\sqrt{\frac{3}{8\pi G}\Omega_Q(1+w_0)}\nonumber\\
&&\times\frac{4a\Psi'+3\Psi}{a^2}=0\label{e2phi0}
\end{eqnarray}
\end{widetext}
for $c^2_a=0$.

Using their solutions (four fundamental) for $\Psi(k,a)$ and $Q(k,a)$ and the constraint equation
\begin{equation}
 -k^2\Psi=4\pi Ga^2\left(\bar{\rho}_M D^{(M)}+\bar{\rho}_Q D^{(Q)}\right),\label{constrainteq}
\end{equation}
it is possible to find the values of $D_g^{(M)}(k,a)$, $V^{(M)}(k,a)$, $D_g^{(Q)}(k,a)$, $V^{(Q)}(k,a)$ and $\Gamma^{(Q)}(k,a)$. The equations (\ref{cons_M1})-(\ref{cons_M2})
can be substituted by one second-order equation
\begin{eqnarray}
{D_g''}^{(M)}+(2-q)\frac{{D_g'}^{(M)}}{a}+\frac{k^2}{a^4H^2}\Psi=0.\label{cons_M3}
\end{eqnarray}

The systems of equations (\ref{e2psiw})-(\ref{e2phiw}) and (\ref{e2psi0})-(\ref{e2phi0}) describe the evolution of perturbations of gravitational $\Psi$ and quintessentional $\delta Q$ fields and their coupling. A few important conclusions can be deduced from qualitative analysis of these systems:
\begin{itemize}
\item The coupling of $\Psi-$ and $\delta Q-$field is modulated by the value $\sqrt{\Omega_Q(1+w)}$ for $w=const$ and $\sqrt{\Omega_Q(1+w_0)}$ for $c^2_a=0$.  So, if $w=w_0=-1$ then both fields evolve independently. Since observational data prefer current $w$ close to $-1$, so their coupling is weak.
\item Evolution of $\delta Q-$field depends on relation of scale of perturbation to horizon explicitly while the dependence of $\Psi-$field is implicit (through the latter). 
\item The system of equations for $w$-quintessence (\ref{e2psiw})-(\ref{e2phiw}) allows the asymptotic behaviour $\Psi\rightarrow const$, $\delta Q\rightarrow const$ when $a\rightarrow \infty$ with relation between them $\delta Q=\Psi/\sqrt{6\pi G(1+w)}$. 
\item The system of equations for $c^2_a$-quintessence (\ref{e2psi0})-(\ref{e2phi0}) allows the asymptotic behaviour $\Psi\rightarrow const$, $\delta Q\rightarrow const$ when $a\rightarrow 0$ with initial relation between them $\delta Q=\sqrt{\Omega_Q(1+w_0)/6\pi G(1+\Omega_Qw_0)}\Psi$. 
\item From equation (\ref{cons_M3}) it follows that $D_g^{(M)}\approx const$ for superhorizon perturbations ($k\ll a^2H$). If $\Psi=const$ and $q=1/2$ at $a\rightarrow 0$ then either $D_g^{(M)}=const-2ak^2\Psi/3H_0^2(1-\Omega_Q)$ for $w=const$ or $D_g^{(M)}=const-2ak^2\Psi/3H_0^2(1+\Omega_Qw_0)$ for $c_a^2=0$ and for $\Psi<0$ it begins to grow slowly from the constant value. The decay 
of $\Psi$ and transition from deceleration to acceleration slow the growth of $D_g^{(M)}$.
\end{itemize}

\subsection{Initial conditions}

Analysis of the background dynamics presented in the previous section has shown that both QCDM-models are matter-dominated in the early Universe (Fig. \ref{fig1}). In the QCDM-model with $w=const$ the ratio $\rho_M/\rho_Q\rightarrow\infty$ when $a\rightarrow 0$, while in the QCDM-model with $c^2_a=0$  $\rho_M/\rho_Q\rightarrow (1-\Omega_Q)/(1+w_0)\Omega_Q$ and $w\rightarrow 0$ when $a\rightarrow 0$. The adiabatic growing mode of perturbation in the non-relativistic matter-dominated Universe can be specified by the condition $\Psi=const$ ($\dot\Psi=0$). Adiabaticity condition in two-component model 
($S_{M:Q}\equiv D_g^{(M)}-D_g^{(Q)}/(1+w)=0$ \cite{doran2003}) gives $D_g^{(M)}=D_g^{(Q)}/(1+w)$. These conditions, constraint equations written for hypersurface $\eta_{init}\ll \eta_0$ ($a_{init}\ll 1$) and the analytic asymptotic solutions (see next subsection) lead to the following adiabatic initial conditions:
\begin{eqnarray}
&&{V^{(Q)}}_{init}=\frac{2}{3}\frac{k}{H_0}\frac{\Psi_{init}}{\sqrt{1-\Omega_{Q}}}\sqrt{a_{init}}\label{ivQw},\\
&&{D_g^{(Q)}}_{init}=-5(1+w)\Psi_{init}\label{idgQw},\\
&&{V^{(M)}}_{init}=\frac{2}{3}\frac{k}{H_0}\frac{\Psi_{init}}{\sqrt{1-\Omega_{Q}}}\sqrt{a_{init}}\label{ivMw},\\
&&{D_g^{(M)}}_{init}=-5\Psi_{init}\label{idgMw}
\end{eqnarray}
for $w=const$ and 
\begin{eqnarray}
&&{V^{(Q)}}_{init}=\frac{2}{3}\frac{k}{H_0}\frac{\Psi_{init}}{\sqrt{1+\Omega_{Q}w_0}}\sqrt{a_{init}}\label{ivQ0},\\
&&{D_g^{(Q)}}_{init}=-5\Psi_{init},\label{idgQ0}\\
&&{V^{(M)}}_{init}=\frac{2}{3}\frac{k}{H_0}\frac{\Psi_{init}}{\sqrt{1+\Omega_{Q}w_0}}\sqrt{a_{init}}\label{ivM0},\\
&&{D_g^{(M)}}_{init}=-5\Psi_{init}\label{idgM0}
\end{eqnarray}
for $c^2_a=0$.

Therefore, the growing mode of adiabatic perturbations in two-component (non-relativistic matter and quintessence) medium is defined by single value -- initial gravitational potential $\Psi_{init}$.

Since the non-adiabatic initial perturbations are strongly constrained by WMAP data, in this paper we restrict ourselves only to adiabatic initial conditions.

\subsection{Asymptotic and numerical solutions}

\begin{figure*}
\includegraphics[width=16cm]{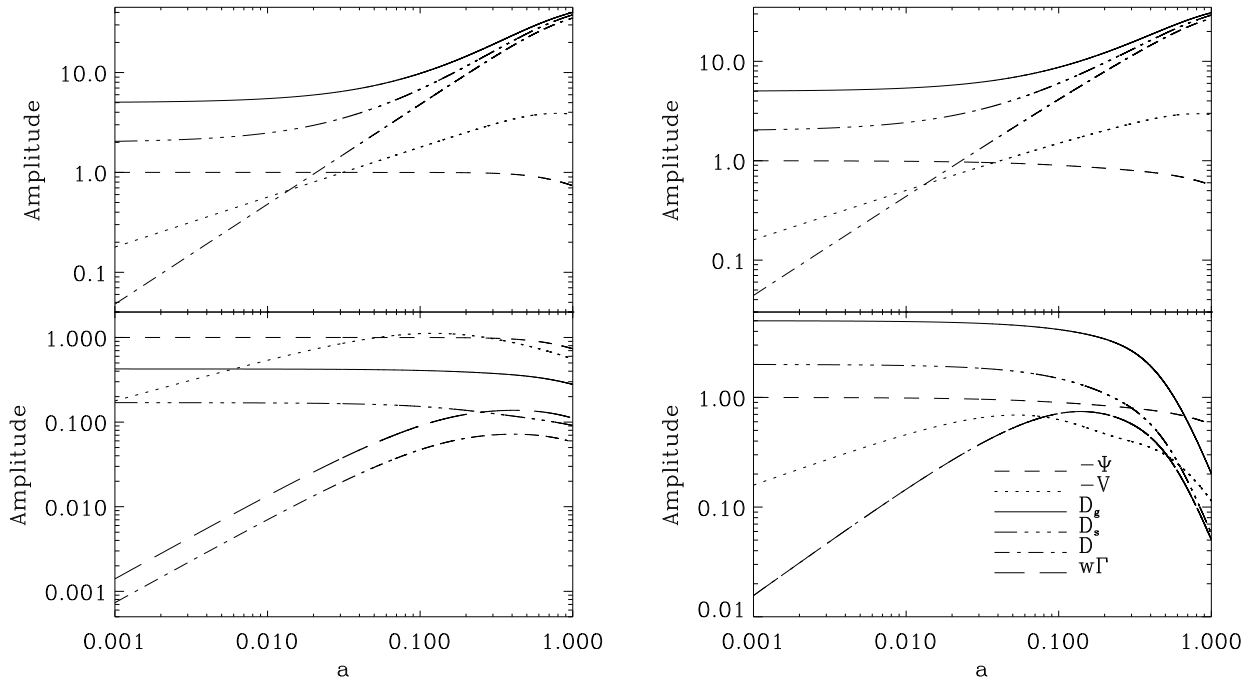}
\caption{The evolution of gauge-invariant amplitudes of perturbations in matter (top) and quintessence (bottom)  for two models of quintessence: $w=const$ (left column) and $c^2_a=0$ (right column). The corresponding scale of perturbations is $k=0.001Mpc^{-1}$ and the cosmological parameters are $\Omega_{Q}=0.745$, $w=-0.915$, $\Omega_M=0.255$, $h=0.7$.}
\label{fig4}
\end{figure*}
In order to analyse the evolution of gauge-invariant variables of matter and quintessence perturbations we must solve the system of equations (\ref{edgQw})-(\ref{epsiw}) together with (\ref{cons_M1})-(\ref{cons_M2}) numerically for initial conditions (\ref{ivQw})-(\ref{idgMw}) or (\ref{ivQ0})-(\ref{idgM0}) respectively. But before we propose the analysis of these systems of equations in the early
epoch ($a\ll 1$), for which the analytical solutions are known. So, the system of equations (\ref{e2psi0})-(\ref{e2phi0}) for $a\ll 1$ can be 
simplified as 
\begin{eqnarray}
&&\Psi''+\frac{7}{2}\frac{\Psi'}{a}+\frac{3}{2}\frac{\Omega_Q(1+w_0)}{1+\Omega_Qw_0}\frac{\Psi}{a^2}- \sqrt{\frac{6\pi G\Omega_Q(1+w_0)}{1+\Omega_Qw_0}}\nonumber\\
&&\times\frac{2a\delta Q'+3\delta Q}{2a^2}=0, \label{e2psi00}\\
&&\delta Q''+\frac{5}{2}\frac{\delta Q'}{a}+\frac{9}{2a^2}\delta Q-\sqrt{\frac{3}{8\pi G}\frac{\Omega_Q(1+w_0)}{1+\Omega_Qw_0}}\nonumber\\
&&\times\frac{4a\Psi'+3\Psi}{a^2}=0.\label{e2phi00}
\end{eqnarray}
This system of equations has 4 fundamental solutions, so it is possible to write the general solution in the form: 
\begin{widetext}
\begin{eqnarray*}
\Psi&=&C_1+\frac{C_2}{a^{\frac{5}{2}}}+C_3a^{-\frac{3}{4}\left(1+\sqrt{\frac{\Omega_Qw_0+8\Omega_Q-7}{1+\Omega_Qw_0}}\right)}+C_4a^{-\frac{3}{4}\left(1-\sqrt{\frac{\Omega_Qw_0+8\Omega_Q-7}{1+\Omega_Qw_0}}\right)},\\
\delta Q&=&\frac{1}{\sqrt{6\pi G}}\sqrt{\frac{\Omega_Q(1+w_0)}{1+\Omega_Qw_0}}\left[C_1+\frac{3}{2}\frac{C_2}{a^{\frac{5}{2}}}-\frac{1+\Omega_Qw_0}{\Omega_Q(1+w_0)}\sqrt{\frac{\Omega_Qw_0+8\Omega_Q-7}{1+\Omega_Qw_0}}\times\right. \nonumber\\
 &&\left.\left(C_3a^{-\frac{3}{4}\left(1+\sqrt{\frac{\Omega_Qw_0+8\Omega_Q-7}{1+\Omega_Qw_0}}\right)}-C_4a^{-\frac{3}{4}\left(1-\sqrt{\frac{\Omega_Qw_0+8\Omega_Q-7}{1+\Omega_Qw_0}}\right)}\right)\right].
\end{eqnarray*}
\end{widetext}
The first two solutions, noted by the constants of integration $C_1$ and $C_2$, are well known growing and decaying modes of adiabatic perturbations in the dust matter-dominated Universe. The next two solutions, noted by the constants of integration $C_3$ and $C_4$, are due to possible entropy initial conditions and intrinsic non-vanishing entropy of quintessence. Really, the condition $\Gamma^{(Q)}=0$ leads to 1 second-order equation which has two dust-like fundamental solutions: 
\begin{eqnarray*}
\Psi=\tilde{C}_1+\tilde{C}_2a^{-\frac{5}{2}}, \hskip1cm
D_s^{(Q)}=-2\left(\tilde{C}_1-\frac{3}{2}\tilde{C}_2a^{-\frac{5}{2}}\right).
\end{eqnarray*}
For the quintessence with $w=const$  solutions for $\Psi$ are the same and 
$D_s^{(Q)}=-2(1+w)\left(\tilde{C}_1-\frac{3}{2}\tilde{C}_2a^{-\frac{5}{2}}\right)$.

The quantities $D_g^{(M)}(k,a)$, $V^{(M)}(k,a)$ and $V^{(Q)}(k,a)$ can be found using the equations (\ref{ddgds})-(\ref{cons_M2}) and (\ref{constrainteq}). The relations between them are presented in the previous subsection as the set of initial data (\ref{ivQw})-(\ref{idgM0}). 

We have integrated numerically the systems of equations (\ref{edgQw})-(\ref{epsiw}) for $w=const$ and for $c^2_a=0$ together with (\ref{cons_M1})-(\ref{cons_M2}) for adiabatic initial conditions (\ref{ivQw})-(\ref{idgMw}) and (\ref{ivQ0})-(\ref{idgM0}) using the publicly available code DVERK\footnote[1]{It was created by T.E. Hull, W.H.Enright, K.R. Jackson in 1976 and is available at http://www.cs.toronto.edu/NA/dverk.f.gz}. We assumed $a_{init}=10^{-10}$ and integrated up to $a=1$. The evolution of perturbations is scale dependent, so we performed calculations for $k=0.0001,\,\,0.001,\,\,0.01$ and $0.1$ $Mpc^{-1}$ for the cosmological model with parameters $\Omega_{Q}=0.745$, $w=-0.915$, $\Omega_M=0.255$, $h=0.7$. The evolution of gauge-invariant variables of matter perturbations $D_g^{(M)}$, $D_s^{(M)}$, $D^{(M)}$, $V^{(M)}$ for two scales $k=0.001$ and $0.01$ $Mpc^{-1}$ is shown in top panels of Fig.\ref{fig4} and Fig.\ref{fig5}. In the bottom panels the analogical gauge-invariant variables of quintessence perturbations ($D_g^{(Q)}$, $D_s^{(Q)}$, $D^{(Q)}$, $V^{(Q)}$, $w\Gamma^{(Q)}$)  are presented (for $c^2_a=0$ the curves $w\Gamma^{(Q)}$ and $D^{(Q)}$ overlap). The evolution of gauge-invariant gravitational potential $\Psi$ is shown in all panels for comparison. All plots are shown for the following range: $0.001\le a\le 1$.

\section{Discussion}
\begin{figure*}
\includegraphics[width=16cm]{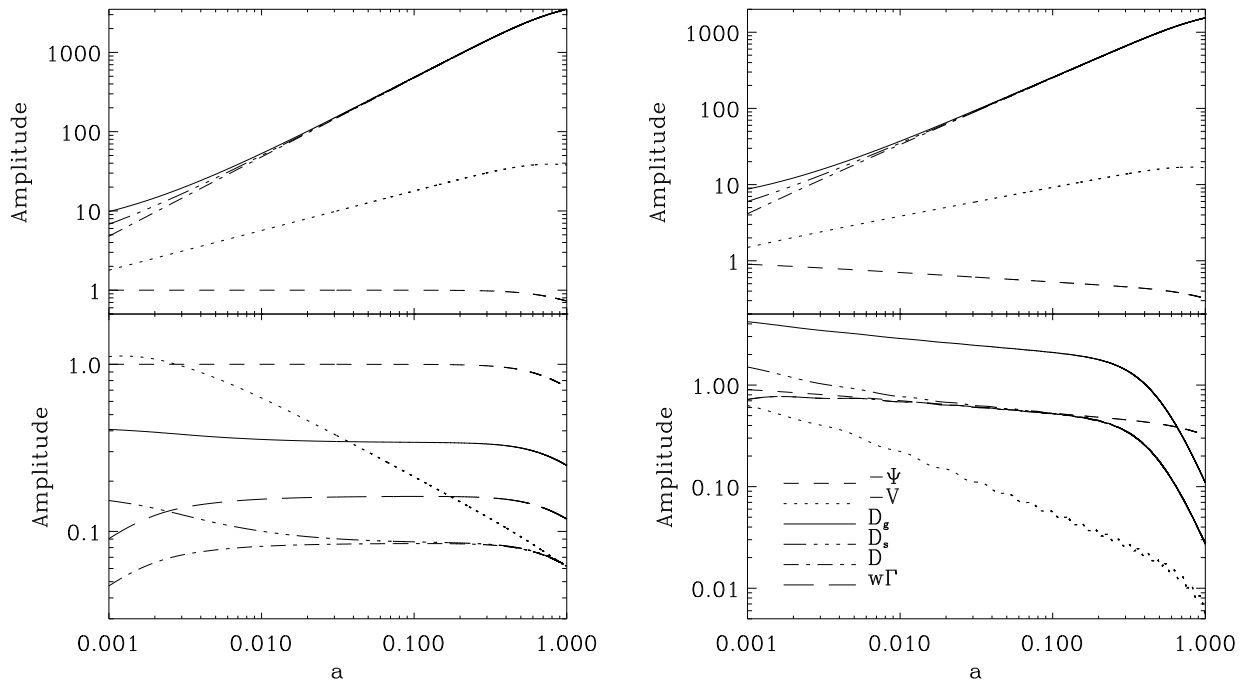}
\caption{The same as in Fig.\ref{fig4} for scale $k=0.01Mpc^{-1}$.}
\label{fig5}
\end{figure*}
\begin{figure}
\includegraphics[width=8cm]{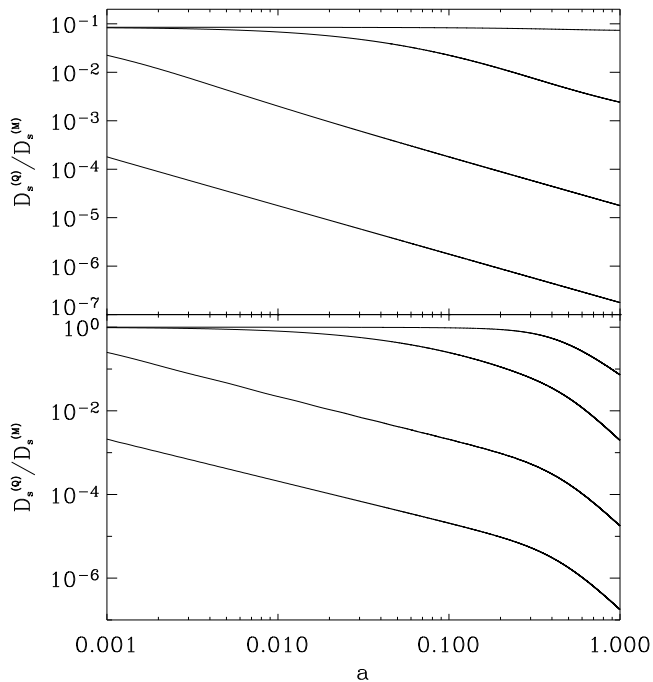}
\caption{Dependences of ratios $D_s^{(Q)}/D_s^{(M)}$ on scale factor for linear perturbations with scales $k=0.0001$, $0.001$, $0.01$ and $0.1$ $Mpc^{-1}$ (from top to bottom) in the models with non-relativistic matter and quintessence ($w=const$ -- top panel, $c^2_a=0$ -- bottom).}
\label{fig6}
\end{figure}

From the top panels of Fig.\ref{fig4}, \ref{fig5} it follows that the magnitudes of the adiabatic matter density perturbations grow similarly in both models (and like in $\Lambda$CDM-model), but time dependences of magnitudes of the adiabatic quintessence energy density perturbations are more varied (bottom panels of the same figures): gauge-invariant variables $D_g^{(Q)}$ and $D_s^{(Q)}$ decay from initial constant value after the particle horizon entry while $D^{(Q)}$ and $V^{(Q)}$ grow at early stage before the horizon entry and decay after that -- in the quintessence-dominated epoch, when gravitational potential starts to decay. The perturbation shown in Fig.\ref{fig4} enters the particle horizon ($\eta(a)=\pi/k$) at $a\approx 0.03$ for $w$-quintessence and at $a\approx 0.04$ for $c^2_a$-quintessence. The perturbation shown in Fig.\ref{fig5} enters the particle horizon at $a\approx 0.0004$ and $a\approx 0.0005$ for $w$- and $c^2_a$-quintessence respectively. The particle horizon at current epoch ($\eta_0$) in the cosmological model with parameters $\Omega_{Q}=0.745$, $w=-0.915$, $\Omega_M=0.255$, $h=0.7$ and $w$-quintessence equals $\approx 14970$ Mpc. In the model with $c^2_a$-quintessence it is $\approx 13810$ Mpc. At early epoch $D^{(Q)}\propto a$ for both models of quintessence. After the particle horizon entry the amplitudes start to decay slowly in the matter-dominated epoch and decay fast in the quintessence-dominated one. At asymptotic regime for quintessence model with $c^2_a=0$ approximately $D^{(Q)}\propto a^{-3}$. In the quintessence model with $w=const$ the transition epoch is extended in time. In Fig.\ref{fig6} we show the dependences of ratios of quintessence density perturbations to matter density ones in conformal-Newtonian gauge ($D_s^{(Q)}/D_s^{(M)}$) on scale factor for perturbations with the scales $k=0.0001$, $0.001$, $0.01$ and $0.1$ $Mpc^{-1}$. These curves emphasise the difference of evolution of perturbations in ordinary matter and quintessence as well as the similarity of behaviour of perturbations in two models of quintessence. The magnitudes of quintessence density perturbations in units of matter ones in both models at current epoch are close although their initial magnitudes differ by order. The magnitudes of quintessence density perturbations with scale less than particle horizon are lower than corresponding magnitudes of matter density perturbations by factor $\approx(23000k)^2$ so, that for scale $k=0.01$  $D_s^{(Q)}/D_s^{(M)}\approx 2\times10^{-5}$. Therefore, on subhorizon scales the quintessential scalar fields are practically smoothed out while the matter clusters.

The tests for choice of the type of dark energy are based on the results of its action on luminous matter and cosmic microwave background. So, the key question is how these types of quintessence affect the growth of matter density perturbations and the time variation of gravitational potential. From top
panels of Fig.\ref{fig4} and \ref{fig5} we can see that they are more suppressed for $c^2_a=0$ than for $w=const$ and for perturbations with smaller scale. In order to illustrate this effect in Fig.\ref{fig7} we present ratios $D^{(M)}a_{init}/D^{(M)}_{init}a$ and $\Psi/\Psi_{init}$ for scales $k=0.0001$, $0.001$, $0.01$ and $0.1$ $Mpc^{-1}$. We can see that scale dependence of suppression of magnitude of matter density perturbations as well as of gravitational potential is strong for $c^2_a=0$ quintessence and weak for $w=const$ one. In the $\Lambda$CDM-model it is scale-independent  \cite{carroll1992}. (In the Einstein -- de Sitter model both ratios are equal to 1 for all times and scales). These ratios are substantial for calculations of magnitude of the matter density power spectrum at different redshifts and the angular power spectrum of CMB temperature fluctuations in the range of scales of the late integrated Sachs-Wolfe effect.

Evolution of quintessence perturbations depends on scalar field model (i.e. its Lagrangian and potential), contents of the Universe, coupling of the quintessence to other components, initial conditions and scale of perturbations \cite{dave2002,doran2003,dedeo2003,bartolo2004,bean2004,liu2004,unnikrishnan2008}. Here we have analysed the evolution of scalar matter and quintessence perturbations for potentials of scalar fields with classical Lagrangian constructed to give either $w=const$ or $c^2_a=0$.  Therefore, obtained here results could be compared to the results of other authors only qualitatively. Evolution of EoS parameter in our $c^2_a=0$-model (Fig.\ref{fig1}) is similar to that of \cite{doran2003}. Despite the other cosmological model and potential of scalar field, the qualitative behaviour of quintessence perturbations is close to obtained here: $D_g^{(Q)}$ is $const$ when the perturbation is outside the particle horizon and decays when it enters the horizon. The growth of magnitude of quintessence density perturbations long before the horizon entry in synchronous gauge was shown by \cite{dave2002} (models with $w=const$ in Fig.3). Our results for evolution of gauge-invariant variable $D^{(Q)}$ (density perturbation in synchronous gauge) shown in Fig.\ref{fig4} support this conclusion. (We do not discuss the oscilations at early stage visible in Fig.3 of \cite{dave2002} because of different initial conditions and background.) The ratios of quintessence ($w=const$, $c^2_s=1$) density perturbations to matter density ones in synchronous gauge are shown in Fig.1 of \cite{bean2004} for $k=0.01$h$^{-1}$ $Mpc^{-1}$. Presented here in Fig.\ref{fig6} analogical ratios in conformal-Newtonian gauge are similar. The conclusion about anti-correlation between the perturbations of the matter and quintessence has been done by \cite{dutta2007} and \cite{mota2008} on the base of analysis of their evolution in the matter rest frame. \cite{bean2004} and \cite{weller2003} noted this effect too. Recalculation of the frame-dependent variables to gauge-invariant ones will -- in our belief -- remove such variance.

\section{Conclusion}
\begin{figure}
\includegraphics[width=8cm]{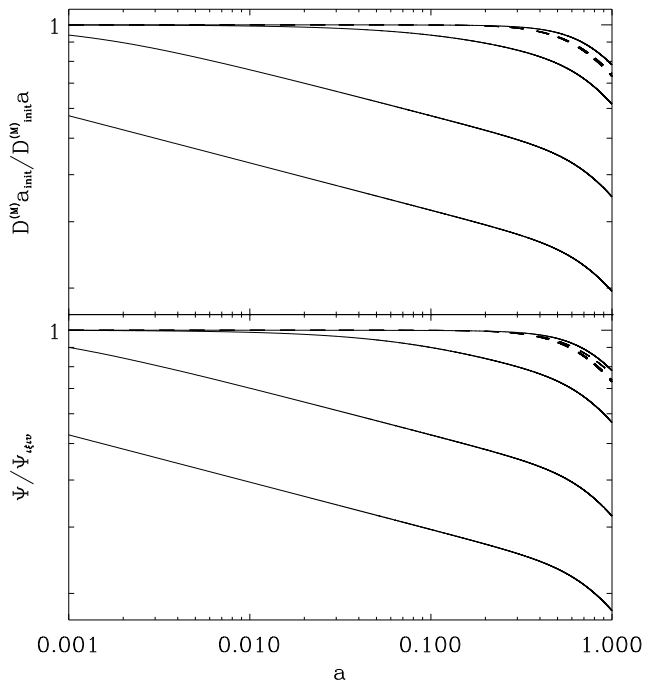}
\caption{Evolution of ratios $D^{(M)}a_{init}/D^{(M)}_{init}a$ and $\Psi/\Psi_{init}$ for linear perturbations with scales $k=0.0001$, $0.001$, $0.01$ and $0.1$ $Mpc^{-1}$ (from top to bottom) in the models with non-relativistic matter and quintessence ($c^2_a=0$ -- solid line, $w=const$ -- dashed line).}
\label{fig7}
\end{figure}
The dynamics of expansion of the Universe and evolution of scalar perturbations are studied for the quintessential scalar fields $Q$ with the classical Lagrangian $L=\frac{1}{2}Q_{;i}Q^{;i}-U(Q)$ satisfying the additional condition $w=const$ or $c^2_a=0$. For both quintessential scalar fields the potential $U(Q)$ and time dependence of $Q$ are constructed for the same cosmological model and it is shown that the accelerated expansion of the Universe is caused by the effect of rolling down of the potential to minimum (Fig.\ref{fig3}). In QCDM-model with $w=const$ the ratio $\rho_M/\rho_Q\rightarrow\infty$ when $a\rightarrow 0$, while in QCDM-model with $c^2_a=0$  $\rho_M/\rho_Q\rightarrow (1-\Omega_Q)/(1+w_0)\Omega_Q$ and $w\rightarrow 0$ when $a\rightarrow 0$.
At the early epoch $w$-quintessence is dynamically unsubstantial like cosmological constant while $c^2_a$-quintessence mimics dust matter ($w\approx 0$) at $a\ll 1$ and cosmological constant ($w=-1$) at $a\gg 1$. The dependence of acceleration parameter on redshift is a bit different for them (Fig.\ref{fig1}) but close to $\Lambda$CDM-model one and indistinquishable observationally now.

Asymptotic analysis of the systems of evolutionary equations for gauge-invariant perturbations has shown that adiabatic initial conditions for non-relativistic matter and $w$- and $c^2_a$-quintessence are allowed.  The numerical integration of these systems give time dependences of gauge-invariant variables for matter and quintessence scalar perturbations (Fig.\ref{fig4}, \ref{fig5}). The main conclusion deduced from them is following: the magnitudes of the adiabatic matter density perturbations grow like in $\Lambda$CDM-model, while for quintessence $D_g^{(Q)}$, $D_s^{(Q)}$ are constant and $D^{(Q)}$, $V^{(Q)}$ grow before the particle horizon entry but all variables decay after that in such way, that at the current epoch they are approximately two orders lower than the corresponding quantities for dust matter on supercluster scales. Therefore, on subhorizon scales the quintessential scalar field is smoothed out while the matter is clustered.

The quintessential scalar fields studied here suppress the growth of matter density perturbations and the  magnitude of gravitational potential (Fig.\ref{fig7}). In these QCDM-models -- unlike $\Lambda$CDM ones -- such suppression is scale dependent and more visible for $c^2_a$-quintessence.
Such features of quintessence are important for calculations of the matter density power spectrum at different redshifts and the power spectrum of CMB temperature fluctuations in the range of scales of the late integrated Sachs-Wolfe effect. That can be used for interpretation of data of current and planned experiments in order to identificate the nature of dark energy.

\begin{acknowledgements}
This work was supported by the project of Ministry of Education and Science of Ukraine ``The linear and non-linear stages of evolution of the cosmological perturbations in models of the multicomponent Universe with dark energy'' (state registration number 0107U002062) and the research program of National Academy of Sciences of Ukraine ``The exploration of the structure and components of the Universe, hidden mass and dark energy (Cosmomicrophysics)'' (state registration number 0107U007279). The authors are thankful also to Yu. Kulinich for useful discussions.
\end{acknowledgements}

\end{document}